\documentclass[twocolumn,showpacs,preprintnumbers,amsmath,amssymb]{revtex4}

\usepackage{graphicx}
\usepackage{bm}
\usepackage{amsmath}
\usepackage{amsfonts}

\begin{document}
\title{Optical tweezers for vortex rings in Bose-Einstein condensates}
\date{\today}
\author{A.I. Yakimenko$^{1}$, Yu.M. Bidasyuk$^2$, O.O. Prikhodko$^1$, S.I. Vilchinskii$^1$,
E.A. Ostrovskaya$^{3}$, and Yu.S. Kivshar$^{3}$
}
\affiliation{$^1$Department of Physics, Taras Shevchenko National University, Kyiv 01601, Ukraine \\
$^2$Bogolubov Institute for Theoretical Physics, National Academy of Ukraine, Kyiv 03680, Ukraine\\
$^3$Nonlinear Physics Centre, Research School of Physics and Engineering, The Australian National University, Canberra ACT
0200, Australia}

\begin{abstract}
We study formation and stabilization of vortex rings in atomic Bose-Einstein condensates. We suggest a novel
approach for generating and trapping of vortex rings by 'optical tweezers'--two blue-detuned optical beams forming a toroidal void in the bulk of a magnetically or optically confined condensate cloud. We demonstrate that matter-wave vortex rings trapped
within the void are energetically and dynamically stable.  Our theoretical findings suggest the possibility for generation, stabilization, and nondestructive manipulation of quantized vortex rings in experimentally feasible trapping configurations.
\end{abstract}

\pacs{05.30.Jp, 03.75.Kk, 03.75.Lm}

\maketitle

A vortex ring is one of the most fascinating and universal structures in fluids of different nature.
Well-known examples are smoke rings produced by smokers or an active volcano, bubble rings created by dolphins, and vortex rings in the blood stream produced  by human heart. Vortex rings have been the subject of numerous studies in classical fluid mechanics \cite{Saffman}.



In quantum fluids and degenerate gases, vortex rings with a closed-loop core occupy a special place among other nonlinear excitations.
Vortex rings play a crucial role in any decay of superflow and in quantum turbulence \cite{Tsubota13}.
Several experimental schemes for creating vortex rings in atomic Bose-Einstein condensates (BECs), based on dynamical instabilities of collective excitations \cite{PhysRevLett.86.2926} or condensate collisions \cite{Shomroni2009,Hau2005}, have been successfully tested. Additional theoretical proposals involve collisions of two-component BECs with different velocities \cite{Jackson99},
 space-dependent Feshbach resonance \cite{Berloff13},  or phase imprinting methods \cite{Ruostekoski01}.

Despite the multitude of methods to generate vortex rings in inhomogeneous trapped BECs, they turn out
to be unstable (see e.g. \cite{Jackson99}), which substantially restricts their lifetime and complicates
experimental observation. In practice, the vortex ring either drifts to an edge of the condensate,
where it immediately decays into elementary excitations, or shrinks and annihilates within the condensate bulk. To the best of our knowledge, a stable vortex ring in a trapped condensate has not been demonstrated either theoretically or experimentally.

In this paper, we propose an experimentally
feasible trapping configuration that can be used to create, stabilize, and manipulate a vortex ring in a controllable and nondestructive
manner.  Our method for the vortex-ring stabilization is based on
a simple physical observation: when a superfluid flow involves fewer atoms, the energy cost
to nucleate a vortex ring decreases because of a smaller contribution to the kinetic energy of the
superfluid. Thus, the spatial position of the vortex core in a toroidal 'anti-trap' with the
locally depressed atomic density is energetically preferable. In contrast to such anti-trapping configuration, a vortex ring in a toroidal trap was found to be unstable \cite{Pi08}. Here we propose to use repulsive
blue-detuned laser beams to create a toroidal void in the bulk of the BEC cloud held in a large-scale magnetic or optical trap, which can be used to trap and guide a vortex ring. We demonstrate both energetic and dynamical stability of the vortex rings for realistic experimental parameters.

\begin{figure}
\includegraphics[width=3.4in]{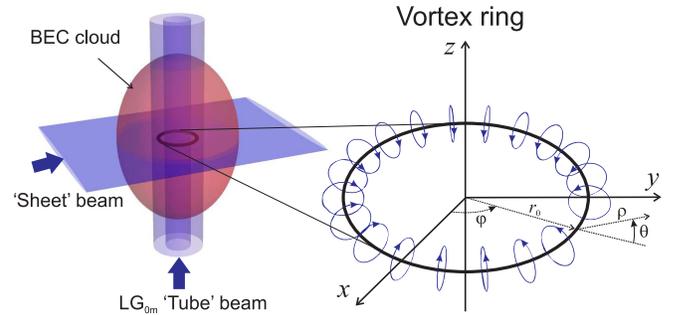}
\caption{(Color online) Left: The proposed trapping scheme for creation and manipulation of vortex rings. Right:  Schematics of the condensate flow in a vortex ring. The circulation directions of the super-flows swirling around the closed-loop core (black circle)  are indicated by arrows. 
}\label{Ring}
\end{figure}

\textit{Model} -- Dynamical properties of an ultracold dilute atomic BEC can be accurately described by the mean-field Gross-Pitaevskii  equation (GPE):
\begin{equation}\label{GPE}
i \hbar \frac{\partial \tilde\Psi(\textbf{r},t)}{\partial t} = \left[-\frac{\hbar^2}{2M} \Delta + \tilde V(\textbf{r}) + U_0 |\tilde\Psi(\textbf{r},t)|^2 \right]\tilde\Psi(\textbf{r},t),
\end{equation}
where $\Delta$
is a Laplacian operator, $U_0 = \frac{4 \pi \hbar^2 a_s}{M}$ is coupling strength, $M$ is the mass of the atom, $a_s$ is the $s$-wave scattering length. The norm of the condensate wave function is equivalent to the number of atoms:
\begin{equation}\label{N}
N=\int|\tilde \Psi|^2d\textbf{r}.
\end{equation}
 Both the number of atoms and the energy,
\begin{equation}\label{E3D}
\tilde E=\int\left\{ \frac{\hbar^2}{2M} |\nabla\tilde\Psi|^2  +\tilde V_\textrm{ext}(\textbf{r})|\tilde\Psi|^2+\frac{U_0}{2}|\tilde\Psi|^4\right\}d\textbf{r},
\end{equation}
are the integrals of motion of Eq. (\ref{GPE}).

A vortex ring manifests itself in emergence of a toroidal hole in the condensate, around which the atoms rotate poloidally with
velocities $\textbf{v}_s$ subject to the condition of quantized circulation (see, e.g., \cite{Kao07}):
\begin{equation}\label{circulation}
\frac{M}{2\pi}\oint_\Gamma \widetilde{\textbf{v}}_sd\textbf{l}=\hbar S,
\end{equation}
where $\Gamma$ indicates a closed contour around the vortex core, the integer $S$ is topological charge of the vortex ring, so that the circulation $Q=2\pi\hbar S/M$ is quantized.
The radius of the vortex core, $a_c$, is usually approximated by
the healing length $a_c \sim \tilde\xi =1/\sqrt{8\pi \tilde na_s}$, where $\tilde n$ is the local condensate density
in a zero-vorticity state.

   The phase, $\Phi$, of the BEC order parameter $\tilde\Psi=|\tilde\Psi|e^{i\Phi}$ with the vortex ring is determined as follows:
   $\Phi(r,z)\approx S\,\theta$, where $r=\sqrt{x^2+y^2}$, which corresponds to a swirling superflow velocity
   $\widetilde{\textbf{v}}_s=(\hbar/M)\nabla\Phi\approx   \textbf{e}_\theta \hbar S /( M l_z \rho)$,
where  $\rho$ is the dimensionless distance from the ring core with the cylindrical coordinates $(r_0,z_0)$,
$\rho=\sqrt{(r-r_0)^2+(z-z_0)^2},$ and $\theta$ is the
poloidal angle  $\tan\theta(r,z)=(z-z_0)/(r-r_0)$  (see Fig. \ref{Ring}).
In what follows we use the dimensionless units for spatial coordinates $(x,y,z)\to (x/l_z,y/l_z,z/l_z)$, where
$l_z=\sqrt{\hbar/(M\omega_z)}$ is the longitudinal  oscillatory length, for time $\tau=\omega_z t$, and for the wave function $\Psi= \tilde\Psi\sqrt{l_z^3}$.
Dimensionless values are denoted by symbols without tilde:
 $E=\tilde E/(\hbar\omega_z)$, $\xi=\tilde\xi/l_z$, $n=|\tilde \Psi|^2\,l_z^3$, $\textbf{v}_s=\widetilde{\textbf{v}}_s/(l_z\omega_z)$, etc.

In our model, the trapping potential is created by a large-scale spherically symmetric harmonic (magnetic) trap and two blue-detuned laser beams, namely a radial Laguerre-Gaussian beam \cite{LGtrap} and
an elliptic highly anisotropic 'sheet' beam, creating a tight repulsive potential in $z$-dimension. The combined potential is given by
\begin{equation}\label{potential}
V=\frac12(z^2+r^2)+v_m(\beta \, r)^{2m}e^{-m\left[(\beta\, r)^2-1\right]}+v_ze^{-\alpha^2 z^2},
\end{equation}
 where
 $\alpha=l_z/Z_\textrm{T}$, $\beta=l_z/R_{\textrm{T}}$, $Z_\textrm{T}$ is the effective width of the sheet-beam, $R_\textrm{T}$ is the radial coordinate of the trap minima, $m$ is the topological charge of the red-detuned LG$_{0m}$ optical beam. Here we consider single-charge ($S=1$) vortex rings in a  BEC cloud  with $N= 10^6$ of $^{87}$Rb atoms ($a_s=5.77$nm) in a harmonic trap with trapping frequency $\omega_z=100$Hz, and corresponding oscillator length $l_z=5.88\mu$m and a tube beam with parameters $m=1$,  $\beta=0.4$ ($R_\textrm{T}=6.57\mu$m), and $v_m=28$, unless specified otherwise.

In modelling non-equilibrium behavior, such as nucleation of vortex rings, dissipative effects are of crucial importance since they provide the mechanism for damping of elementary excitations in the process of relaxation to an equilibrium state. It is the dissipation that either causes the vortex core to drift to the BEC cloud edge (where vortex rings decay), or leads to the relaxation to the metastable state corresponding to the local minimum of the nucleation energy. 


Such effects naturally arise in a trapped condensate due to interaction with uncondensed atoms and can be captured  phenomenologically by the dissipative GPE (DGPE) derived by Choi et al. \cite{Choi98}, in a manner similar to that originally proposed by Pitaevskii \cite{Pitaevskii59}. For a system close to thermodynamic equilibrium and subject to weak dissipation, the DGPE can be written in the form:
\begin{equation}\label{GPE_dissipative1}
i(1+i\gamma)\frac{\partial \psi}{\partial \tau} = \left[-\frac{1}{2} \Delta + V(r,z) +  g|\psi|^2 -\mu\right]\psi,
\end{equation}
were we introduced the chemical potential of the equilibrium state, $\mu$, as follows: $\Psi(\textbf{r},\tau)=\psi(\textbf{r},\tau) e^{-i\mu\tau}$, $-i\partial \psi/\partial t \approx \mu \psi$, and the $s$-wave scattering is characterized by the rescaled constant $g=4\pi a_s/l_z$. The DGPE in the above form  has
been used extensively in previous studies of vortex dynamics \cite{Tsubota03,Tsubota13,PhysRevA.77.023605}.

The phenomenological damping rate, $\gamma \ll 1$, depends both on the temperature and spatial coordinates, and  can be approximately calculated using quantum kinetic theory \cite{Choi98}. Alternatively, its value could be estimated by fitting the vortex lifetime, obtained from numerical simulations, to the results of experimental studies of the dissipative vortex dynamics. In what follows, we neglect the position dependence of $\gamma$, and set  $\gamma=0.03$ as in Ref. \cite{Tsubota13}, however, our main results do not depend qualitatively on the specific value
of $\gamma \in [10^{-3}, 10^{-1}]$.

Dynamical Eq. (\ref{GPE_dissipative1}) with $\gamma>0$ conserves neither the energy
 nor the number of particles. In our simulations, we adjust $\mu(\tau)$ on every time step so that the  number of condensed particles slowly decays with time: $N(\tau)=N(0)e^{- \tau/\tau_0}$  to match the experimental measurements for atomic BECs. In our calculations we use the value of the parameter $\tau_0=10^3$, which corresponds to $t=10$ s for the $1/e$ lifetime of the BEC reported
  in \cite{PhysRevLett.110.200406}.

\begin{figure}
\includegraphics[width=3.4in]{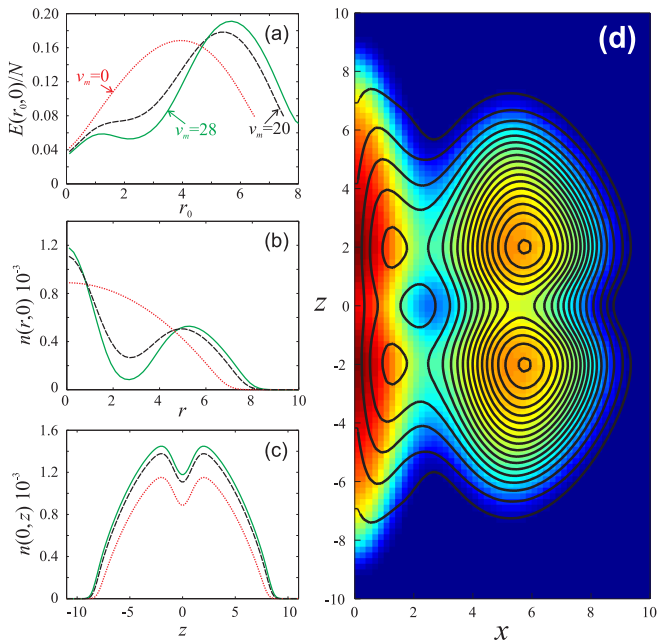}
\caption{(Color online) (a) The nucleation energy of the vortex ring per particle $E(r_0,0)/N$ for different values of the tube-beam intensity. The values of $v_m$ are indicated near the curves; other parameters are: $v_z=10$, $\alpha=0.8$, $\beta=0.4$,  and  $N=10^6$. Density profiles $|\psi|^2\cdot 10^{-3}$ of the condensate in ground state for the same values of $v_m$: (b) The radial part of the density profile  at $z=0$; (c) the longitudinal distribution of the density at $r=0$. (d) Distribution of $|\psi(r,z)|$ with contour lines for nucleation energy.
}\label{EnergeticStability}
\end{figure}

\textit{Energetic and dynamical stability} -- To elucidate stability properties of a vortex ring in the trapping geometry described by Eq. (\ref{potential}), we discuss the energetic stability of the vortex rings in dissipationless condensate and then support our findings by numerical simulation of the DGPE (\ref{GPE_dissipative1}).

In a homogeneous BEC, by the analogy with classical fluids and liquid helium, the  energy  of a large vortex ring can be estimated  as follows (see, e.g. \cite{Barenghi09}):
 $E_n=  2\pi^2 S^2 n r_0\left[\ln(8r_0/a_c)-C_1\right]$,
where $C_1\approx 1.615$,
$r_0$ is the radius of the ring, and $a_c$ is the radius of the vortex core ($r_0\gg a_c$). The Magnus force, which is directed outwards orthogonally to the ring velocity, stabilizes the ring against shrinking. As a result, in homogeneous condensate the ring travels with a constant speed $v_{\textrm{vr}}=\frac{S}{r_0}  \left[\ln(8 r_0/a_c)-C_2\right]$, $C_2\approx 0.615$.

In a trapped, inhomogeneous BEC
the vortex core undergoes precession around an \textit{unstable} equilibrium position, corresponding to the maximum of the nucleation energy \cite{Jackson99}. This is not surprising for a common axisymmetric trap. Indeed, the vortex energy
 depends  on the product of the vortex ring radius $r_0$ and the local value of the density $n$, thus one might expect that the vortex nucleation energy decays both at the edge of the condensate, where density vanishes (close to the Thomas-Fermi surface), and when the vortex ring shrinks towards the axis of the trap where density is nearly constant. Certainly, the inhomogeneity of the condensate density must be properly accounted  for in calculations of nucleation energy, but this  correction, as we show below, does not change the main conclusions qualitatively.

Since the nucleation energy, $E_{\textrm{n}}=E_\textrm{v}-E_\textrm{GS}$,
 is the energy difference for the state with imprinted vortex ring and the ground state,
  the main
  contribution to the nucleation $E_{\textrm{n}}$ energy
  is given by the kinetic energy
  of the poloidal superflows:
$E_{\textrm{n}}\approx \frac12 \int n \textbf{v}_s^2 d\textbf{r}=\frac12 S^2\int n\rho^{-2} d\textbf{r}.$
Let us calculate the vortex ring energy, accounting for the condensate inhomogeneity caused by the trapping potential and toroidal hole
around vortex core.
First, using the imaginary time propagation method, we find
numerically the ground state $\psi_{\textrm{GS}}(r,z)$ of a BEC in
the given trapping potential.
Then we imprint  the off-set vortex ring core onto the ground state:
\begin{equation}\label{ansatz}
\Psi(r,z)=A\left\{f(r,z)\right\}^S\psi_{\textrm{GS}}(r,z)e^{iS \theta},
\end{equation}
where $A$ is the constant of normalization introduced to preserve the number of atoms,  the function $f(r,z)=\tanh\left(\rho/\xi\right)$
interpolates the order parameter $\Psi$ properly both in the vicinity of the core:
$\Psi\sim\rho^S$, if  $\rho\to 0$,  and well away from the core: $|\Psi|\sim|\psi_\textrm{GS}|$,
if  $\rho\gg\xi$, where  $\xi=\sqrt{\frac{l_z}{8\pi a_s}}|\psi_\textrm{GS}(r,z)|^{-1}$
 is the dimensionless healing length.
Using the ansatz (\ref{ansatz}) we obtain the nucleation energy 
 as the function of the vortex core coordinate.
The pronounced minimum of $E_n$ in the vicinity of the minimum of the condensate density is clearly seen from Fig \ref{EnergeticStability} (a), where the nucleation energies per particle are shown for different values of the tube-beam intensitities, $v_m$.

We have verified our energetic stability analysis by extensive series of numerical simulations of DGPE with initial condition obtained by imprinting the off-set vortex ring core onto the ground state. For numerical simulation of Eq. (\ref{GPE_dissipative1}), the split-step Fourier transform method was used. By changing the initial position of the vortex core we have found remarkable agreement between the results of direct numerical simulations and predictions based on the approximate energetic stability analysis. 

\begin{figure}
\includegraphics[width=3.4in]{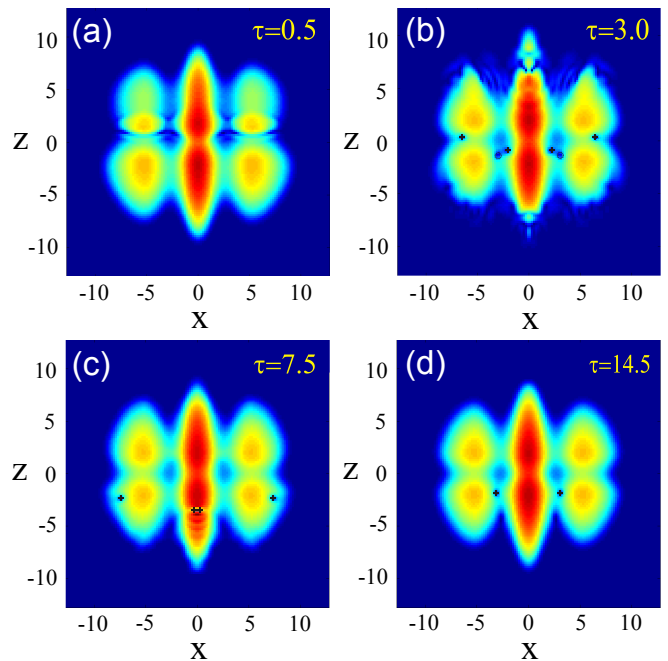}
\caption{(Color online) Typical example of vortex ring generation and further stabilization. The snapshots of the cross-section $|\psi(x,0,z)|$ are shown for different moments of dimensionless time $\tau=\omega_z t$. Parameters of the sheet-beam  change from $\alpha=0.4$, $v_z=20$ and centered at $z=0.5$ to $\alpha=0.8$, $v_z=10$, and centered at $z=0$.}\label{Generation}
\end{figure}
\begin{figure}
\includegraphics[width=3.4in]{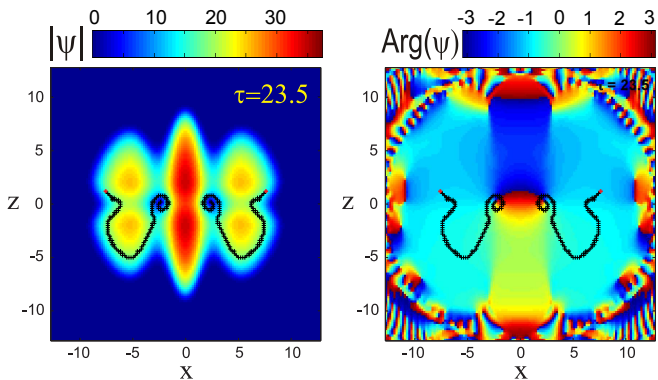}
\caption{(Color online) The trapped and stabilized vortex ring as the result of vortex ring nucleation presented in Fig. \ref{Generation}. Shown are $|\psi|$ (left) and phase (right) of the order parameter at $\tau=23.5$. The subsequent positions of the vortex core of the stable vortex ring are marked by the sequence of black crosses. The red dots indicate the position where the nucleated vortex ring was first detected.
}\label{FinalState}
\end{figure}

\textit{Vortex ring generation} --
Having established energetic and dynamical stability of vortex rings in the toroidal void created by the trapping potential (\ref{potential}), we now analyse two methods for their nucleation, trapping and stabilization. The first method employs interference between BECs loaded in large-scale harmonic potential and separated by a potential barrier. A similar method was used in theoretical \cite{PhysRevA.77.023605} and experimental investigations \cite{Shomroni2009} of the vortex ring generation, where a tight (in $z$-direction)
barrier rapidly vanished, with the resulting interference between two identical condensates giving rise to pairs of the vortex-antivortex rings. Here we use the barrier that initially separates the condensate into two unequal parts, and then is lowered non-adiabatically rather than being completely removed.

A typical example of subsequent evolution displaying generation and trapping of the vortex rings is shown in Fig. \ref{Generation}. As soon as the barrier becomes sufficiently low, both the repulsive interatomic interaction and the trapping potential propel the atoms from the upper half of the condensate to fill in the appearing emptiness. The various vortex rings are generated during collision of the upper half of the BEC cloud with the sheet beam and the rest of the condensate. The conservation of circulation governs nucleation of the two types of vortex rings: (i) single vortex rings appear only
at the boundary of the condensate, (ii) vortex-antivortex rings appears in the bulk
of the condensate and form a moving bound pair, as seen in Fig.  \ref{Generation} (b). The cores of the vortex rings are indicated by crosses, anti-vortex rings are marked by circles. The vortex-antivortex pair moves in the direction of the initial momentum until the anti-vortex approaches the Thomas-Fermi surface of the condensate, where it decays. The remaining  vortex ring comes back to the condensate and annihilates [see Fig. \ref{Generation} (c)]. We stress that the single vortex appears outside the Thomas-Fermi surface of condensate 
then it shrinks and enters the BEC cloud at the depth corresponding to its energy. The nucleated vortex moves close to the surface of constant energy until it reaches the toroidal void [see Fig.  \ref{Generation} (d)] and relaxes to the stable state as seen in Fig. \ref{FinalState}. Thus the single vortex ring, born on the periphery, drifts to the stable position, where it is caught in the toroidal void .

\begin{figure}
\includegraphics[width=3.4in]{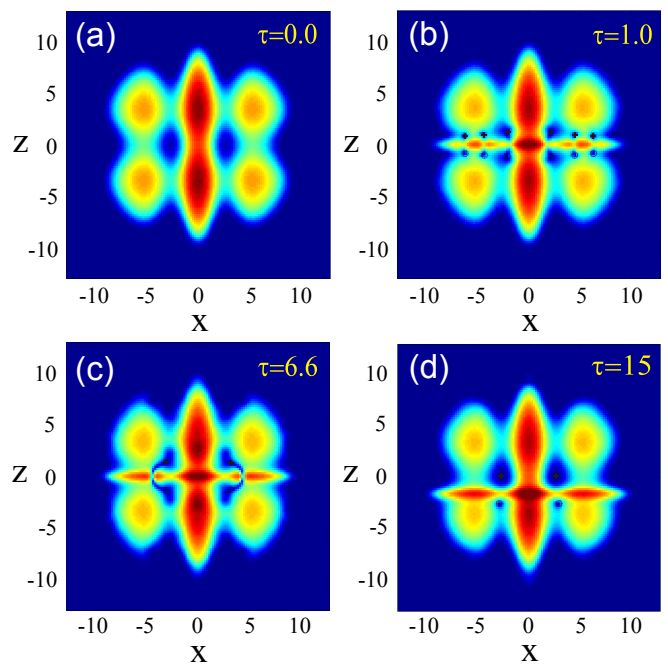}
\caption{(Color online) Generation and stabilization of the vortex-antivortex ring pair with further controllable detachment and destruction of the anti-vortex.
}\label{GenerationPair}
\end{figure}

The second method is designed to stabilize the vortex-antivortex pair inside the condensate, and then to detach and destruct one of vortex rings in a controllable way, so that a single stable vortex ring remains in the final state.
In numerical simulation illustrated in Fig. \ref{GenerationPair}, we  start from a stationary BEC divided in half by a wide blue-detuned repulsive sheet
beam with $\alpha=0.4$, $v_z=20$. The \textit{red-detuned} laser beam creates additional attractive potential $-|v_z^{(r)}(\tau)|e^{-\alpha_r^2z^2}$ with the maximum intensity at $z=0$ and $\alpha_r=2.0$. Starting from zero value,  the intensity of the red-detuned beam ramps up linearly with time so that at $\tau=T_z=0.1$  it reaches the value $|v_z^{(r)}(T_z)|=v_z=20$ and remains unchanged during further evolution. The growing attractive potential draws in the condensed atoms and creates super-critical counter-flows directed to the $z=0$ plane. Conservation of circulation ensures that vortex rings and anti-vortex rings are nucleated simultaneously.  In Fig. \ref{GenerationPair} (b) three vortex-antivortex pair are  seen. Some of the vortex rings move out and decay at the cloud's edge, but one vortex-antivortex pair relaxes to the stationary state, so that each ring is stored in its own toroidal void, separated by the local density maximum at $z=0$. As soon as the pair of vortex rings is stabilized, we start slowly shifting down the red-detuned sheet-beam. When this beam enters and eliminates the lower toroidal void, the anti-vortex ring becomes untrapped. The decoupled anti-vortex ring starts moving to the edge of the cloud and finally decays.  The velocity of the sheet beam, $v_b=0.5$, is chosen to be well below the critical velocity (see e.g. \cite{Raman99,Engels07}), so that new vortex rings appear and decay only in the low-density periphery, i.e. only when the red-detuned beam nearly completely leaves the BEC cloud. After some relaxation time the remaining single vortex ring occupies the stable position. 

Both of the above methods of vortex ring generation and stabilization rely on the formation of a toroidal anti-trap
by a pair of laser beams. The combination of optical trapping and anti-trapping potentials created by these beams represents a kind of \textit{'optical tweezers'} for manipulation of vortex rings; the vortex ring can be 'picked up', stabilized and then moved to a new location by gradually adjusting the parameters of the optical trap.

We note that the novel method for the vortex ring stabilization proposed here  can be embodied in the existing experimental setups. For example, in the very recent experiment \cite{PhysRevLett.110.200406} BEC was trapped in a cylinder potential formed by three blue-detuned laser beams: radial confinement was achieved with a LG tube beam and two sheet beams were used as 'plugs' in the longitudinal direction. 


\textit{Conclusions} --
We have proposed a novel approach for generation and stabilization of vortex rings in the toroidal 'optical tweezers'.
We have proved that the nucleation energy of a vortex ring as the function of the vortex core position has a pronounced minimum within the toroidal void. Strictly speaking, this means that the vortex rings are metastable. In practice, a dissipation of the vortex energy is universally present. As a result, the vortex ring trapped in the region with locally reduced density relaxes to the stationary state with the radius corresponding to a local minimum of the nucleation energy. The lifetime of the vortex ring is restricted only by a decay of the condensate itself.

Our findings not only suggest the possibility for experimental demonstration of the stable vortex rings, but also offer an approach for nondestructive manipulation of vortex rings. We hope that the described method will help to elucidate the properties of the vortex rings in superfluid media, which is the problem of fundamental interest.

The authors thank N. Berloff and C. Conti for useful discussions. The work was partially supported by the Australian Research Council.

\bibliography{BEC_VR}
\end{document}